# Defective high-entropy oxide photocatalyst with high activity for $CO_2$ conversion

Saeid Akrami[1], Yasushi Murakami[2], Monotori Watanabe[3], Tatsumi Ishihara[2,3], Makoto Arita[4], Masayoshi Fuji[1,5,]* and Kaveh Edalati[3,]*

[1] Department of Life Science and Applied Chemistry, Nagoya Institute of Technology, Tajimi 507-0071, Japan
[2] Department of Applied Chemistry, Faculty of Engineering, Kyushu University, Fukuoka 819-0395, Japan
[3] WPI, International Institute for Carbon-Neutral Energy Research (WPI-I2CNER), Kyushu University, Fukuoka 819-0395, Japan
[4] Department of Materials Science and Engineering, Faculty of Engineering, Kyushu University, Fukuoka 819-0395, Japan
[5] Advanced Ceramics Research Center, Nagoya Institute of Technology, Tajimi 507-0071, Japan

**Abstract**

High-entropy oxides (HEOs), as a new family of materials with five or more principal cations, have shown promising properties for various applications. In this work and inspired by inherent defective and strained structure of HEOs, photocatalytic $CO_2$ conversion is examined on a dual-phase $TiZrNbHfTaO_{11}$ synthesized by a two-step high-pressure torsion mechanical alloying and high-temperature oxidation. The HEO, which had various structural defects, showed simultaneous photocatalytic activity for $CO_2$ to CO and $H_2O$ to $H_2$ conversion without the addition of a co-catalyst. The photocatalytic activity of this HEO for $CO_2$ conversion was better than conventional photocatalysts such as anatase $TiO_2$ and $BiVO_4$ and similar to P25 $TiO_2$. The high activity of HEO was discussed in terms of lattice defects, lattice strain, light absorbance, band structure, photocurrent generation and charge carrier mobility to activation centers. The current study confirms the high potential of HEOs as a new family of photocatalysts for $CO_2$ conversion.

*Corresponding authors:
Masayoshi Fuji (E-mail: fuji@ fuji@nitech.ac.jp; Tel: +81-57-227-6811)
Kaveh Edalati (E-mail: kaveh.edalati@kyudai.jp; Tel: +81-92-802-6744)

## 1. Introduction

Carbon dioxide ($CO_2$) is a stable molecule, which is produced mainly from human activities such as combustion of hydrocarbons and partly from natural sources such as decomposition of organisms. Emission of $CO_2$ from combustion of hydrocarbon fuels is considered as a main reason for the global warming on the earth [1,2]. There are currently significant attempts to explore effective strategies to reduce the human-activity-based $CO_2$ emission or to convert $CO_2$ to useful products [1]. Photocatalytic $CO_2$ conversion is considered as a clean technology to reduce the amount of this stable gas in the atmosphere, although the research on photocatalytic $CO_2$ conversion is still at the early stages and significant efforts are required to enhance its efficiency for practical applications [2]. In photocatalytic $CO_2$ conversion, following the light absorbance by a semiconductor which is known as photocatalyst, electrons are excited from the valence band to the conduction band and contribute to $CO_2$ conversion [2]. Such a $CO_2$ conversion occurs through different pathways including carbene pathway, formaldehyde pathway and glyoxal pathway [3,4], leading to different conversion products, as shown in Table 1 [5,6].

Table 1. Reactions in $CO_2$ conversion with their standard potentials [5,6].

| Reaction | Standard Potential vs. NHE (eV) at PH = 0 |
|---|---|
| $CO_2 + 2H^+ + 2e^- \rightarrow HCOOH$ | -0.20 |
| $CO_2 + 2H^+ + 2e^- \rightarrow CO + H_2O$ | -0.11 |
| $CO_2 + 4H^+ + 4e^- \rightarrow HCHO + H_2O$ | -0.07 |
| $CO_2 + 6H^+ + 6e^- \rightarrow CH_3OH + H_2O$ | 0.03 |
| $CO_2 + 8H^+ + 8e^- \rightarrow CH_4 + 2H_2O$ | 0.17 |

For photocatalytic $CO_2$ conversion, a photocatalyst should have appropriate bandgap to absorb light photons, appropriate band structure to satisfy the standard potentials for each reaction shown in Table 1, long exited electron lifetime and appropriate charge carrier mobility [7]. $TiO_2$ [8], $C_3N_4$-based catalysts [9], bismuth-based compounds such as $BiVO_4$ [10], Cu-based semiconductors [11] and $WO_3$ [12] are some popular photocatalysts for $CO_2$ conversion. Although the improvement of existing photocatalysts by different strategies such as heterostructure generation [7], nanosheet production [8], surface defect introduction [9], oxygen vacancy generation [12], mesoporous structure formation [13] and strain engineering [14,15] is currently the major research activity, there are high demands to explore new family of materials as highly active photocatalysts.

High-entropy ceramics are a new type of materials which show high stability and promising structural and functional properties due to the so-called cocktail effect, lattice strain/defects, heterogenous valence electron distribution and high configurational entropy [16]. As shown in Fig. 1a, high-entropy ceramics are defined as multi-component materials with at least five principal elements and a configurational entropy higher than 1.5*R* (*R*: the gas constant) [17]. These materials have a low Gibbs free energy due to their high entropy and this gives a high stability to these materials under different conditions [16] including catalytic reactions [18,19]. Moreover, the presence of at least five cations with different atomic sizes in these materials results in the formation of inherent lattice strain and defects [17]. Since lattice strain and defects are effective to enhance photocatalytic $CO_2$ conversion [9,12,14,15], the high-entropy ceramics are expected to show good activity for such a conversion. High-entropy oxides (HEOs) are the most popular high-entropy ceramics which have been investigated for various applications and properties such as thermal barrier coatings [20,21], magnetic components [22,23], dielectric components [24,25], Li-ion batteries [26,27], Li-S batteries [28], Zn-air batteries [29], catalysts [30,31], electrocatalysts [32], and photocatalytic hydrogen production [33,34]. Despite the inherent defective and strained structure of HEOs, there have been no attempts to employ these materials for photocatalytic $CO_2$ conversion.

In this study, a HEO photocatalyst, $TiZrNbHfTaO_{11}$, is synthesized and its activity for $CO_2$ conversion is examined. The transition elements titanium, zirconium, niobium, hafnium, and tantalum are selected simply because their binary oxides with the $d^0$ electronic structure can act as photocatalysts. This first application of HEOs for photocatalytic $CO_2$ conversion confirms that the HEO photocatalyst shows higher activity compared to common binary or ternary photocatalysts such as $TiO_2$ and $BiVO_4$, suggesting HEOs as a new family of photocatalysts for $CO_2$ conversion.

## 2. Experimental

### *2.1. Sample preparation*

Although various methods have been developed in recent years for the synthesis of HEO [16-34], a two-step high-pressure mechanical alloying and high-temperature oxidation which is available in the authors' laboratory was used to synthesize the HEO. In the first step, equiatomic amounts of Ti (99.9%), Zr (95.0%), Hf (99.5%), Nb (99.9%) and Ta (99.9%) powders were mixed in acetone, treated by ultrasonic and then dried. The dried powder mixture was processed by high-pressure torsion (HPT), shown in Fig. 1b, to fabricate TiZrNbHfTa high entropy alloy (HEA) with the body-centered cubic (BCC) structure (see the principles of HPT and its applications to oxides in [35,36]). To produce the TiZrNbHfTa alloy, a 10 mm diameter and 1 mm thick disc was prepared by compacting the powder mixture under a pressure of 400 MPa. The compacted disc was then compressed between two HPT anvils under a high pressure of 6 GPa at room temperature and simultaneously processed by rotating the lower HPT anvil with respect to the upper one for 100 turns with a rotation rate of one turn per minute. In the second step, the HPT-processed TiZrNbHfTa alloy was exposed to hot air at a temperature of 1373 K for 24 h to produce an oxide, with the appearance shown in Fig. 1c. Examination of the mass of sample before and after oxidation suggested a composition of $TiZrNbHfTaO_{11}$ for the produced oxide. The oxide was crushed after oxidation into the powder form using a mortar and examined by various characterization methods, as described below.

### *2.2. Characterization*

To examine the crystal structure, X-ray diffraction (XRD) using the Cu Kα radiation with a wavelength of $\lambda$ = 0.1542 nm and micro-Raman spectroscopy using a laser source with a wavelength of $\lambda$ = 532 nm were utilized.

Examination of microstructure was conducted by (i) scanning electron microscopy (SEM) with energy dispersive X-ray spectroscopy (EDS) analysis under 15 keV, (ii) transmission electron microscopy (TEM) with selected area electron diffraction (SAED), bright-field (BF) images, dark-field (DF) images, high-resolution images and fast Fourier transform (FFT) analysis under 200 keV, and (iii) scanning-transmission electron microscopy (STEM) with high-angle annular dark-field (HAADF) images and EDS analysis under 200 keV.

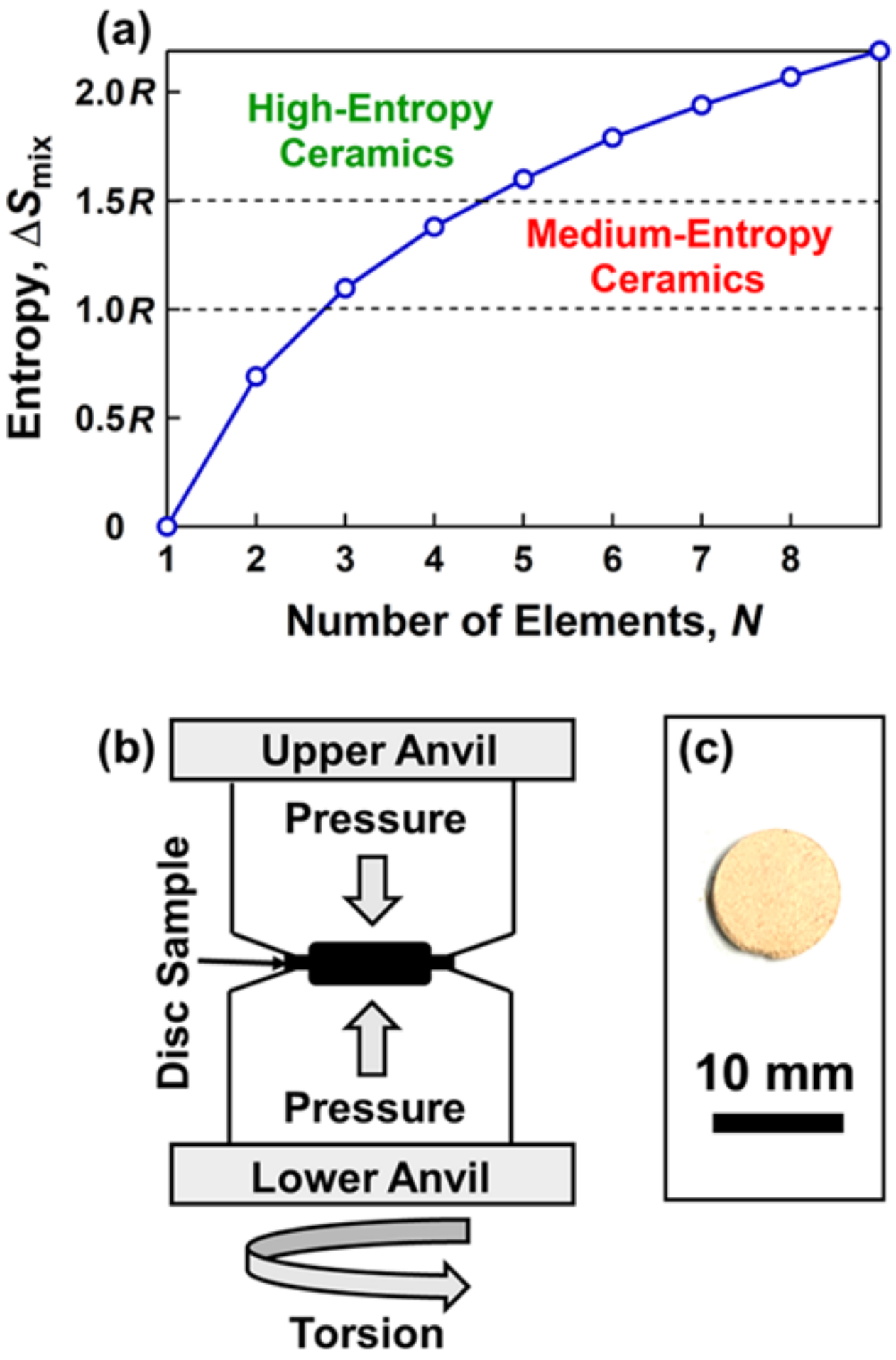

Fig. 1. (a) Relationship between number of elements and configurational entropy and definition of high-entropy ceramics with equiatomic fractions of elements, (b) schematic illustration of high-pressure torsion (HPT), and (c) appearance of high-entropy oxide synthesized in this study.

To investigate the presence of point defects such as oxygen vacancies, electron paramagnetic resonance (EPR) was performed at ambient temperature using a microwave source with a frequency of 9.4688 GHz.

To study the oxidation states of different elements and to estimate the valence band top position, X-ray photoelectron spectroscopy (XPS) using the Al $K_\alpha$ radiation with a wavelength of $\lambda$ = 0.989 nm was used. The XPS energy position for each element was adjusted by considering the peak position of C 1s at 284.8 eV. After correction of the energy positions, the peaks for different elements were analyzed by peak deconvolution by considering the standard energy relations and differences reported in the handbook [37]: $f_{7/2}$:$f_{5/2}$ = 4:3, $d_{5/2}$:$d_{3/2}$ = 3:2, $p_{3/2}$:$p_{1/2}$ = 2:1, Ti $2p_{1/2}$ - Ti $2p_{3/2}$ = 5.54 eV, Zr $3d_{3/2}$ - Zr $3d_{5/2}$ = 2.43 eV, Hf $4f_{5/2}$ - Hf $4f_{7/2}$ = 1.71 eV, Nb $3d_{3/2}$ - Nb $3d_{5/2}$ = 2.72 eV, Ta $4f_{5/2}$ - Ta $4f_{7/2}$ = 1.91 eV.

To investigate the light absorbance and bandgap (Kubelka-Munk analysis), UV-vis diffuse reflectance spectroscopy was conducted, and the band structure was calculated by considering both XPS and UV-vis spectra.

To study the lifetime of exited electrons, steady-state photoluminescence (PL) emission spectroscopy with a 325 nm laser source and time-resolved photoluminescence decay (PL decay) with a 285 nm laser source were conducted.

The specific surface area of powder was examined by nitrogen gas adsorption and using the Brunauer-Emmett-Teller (BET) method.

### *2.3. Photocurrent test*

Photocurrent generation was examined using a thin film of sample in a 1 M $Na_2SO_4$ electrolyte under the full arc of Xe lamp (without using any filter), as described in detail earlier [38]. The thin film was prepared by deposition of HEO powder on FTO (fluorine-doped tin oxide) glass with 2.25 mm thickness and 15×25 $mm^2$ surface area. about 5 mg of sample was crushed in 0.2 mL ethanol and carefully dispersed on the FTO glass using a drop and annealed at 473 K for 24 h. The average thickness of HEO on FTO glass was about 0.04 mm, which was estimated by measuring the thickness of glass before and after deposition of HEO using a micrometer with 0.01 mm accuracy. Photocurrent generation was examined by an electrochemical analyzer in the potentiostatic amperometry mode during time (30 s light ON and 60 s light OFF), while the counter electrode was Pt wire, the reference electrode was Ag/AgCl, and the external potential was 0.7 V vs. Ag/AgCl.

### *2.4. Photocatalytic test*

Photocatalytic $CO_2$ conversion was conducted using the powder of HEO in a continuous flow quartz photoreactor. The photoreactor, as shown in Fig. 2a, had a cylindrical shape with a total inner volume of 858 mL. The reactor had an inner space to insert the light source. There were two holes on the top of photoreactor: one for the inlet of $CO_2$ flow, which was connected to a gas cylinder; and another one for the outlet of gas and sampling the reaction products for analysis, which was connected to a vent and gas chromatograph. For the photocatalytic reaction, 120 mg of HEO was mixed with 500 mL of deionized water and $NaHCO_3$ with 1 M concentration and then bubbled with $CO_2$ with a flow rate of 3 mL/min. The temperature was controlled as 288 K using a water chiller and the suspension was continuously stirred using a magnetic stirrer. The process was first conducted for 2 h without light irradiation, and after confirmation that no reaction products appear, the photocatalytic test was conducted under irradiation with a high-pressure Hg light source (Sen Lights Corporation, HL400BH-8, 400 W, with the spectral composition shown in Fig. 2b). The light intensity irradiated on the photocatalysts was 0.5 $W/cm^2$ and no filter was used during the irradiation. The reaction products were analyzed by a gas chromatograph (Shimadzu GC-8A, Ar Carrier). A flame ionization detector equipped with a methanizer (Shimadzu MTN-1) was used to measure the CO and $CH_4$ production rate. A thermal conductivity detector also was utilized to evaluate the $H_2$ and $O_2$ production. To be sure about the absence of CO from other sources such as contamination, blank tests were conducted (i) under irradiation in the presence of $CO_2$, $NaHCO_3$ and $H_2O$ and without the photocatalyst addition and (ii) under irradiation in the presence of Ar, $NaHCO_3$ and $H_2O$ and with photocatalyst addition.

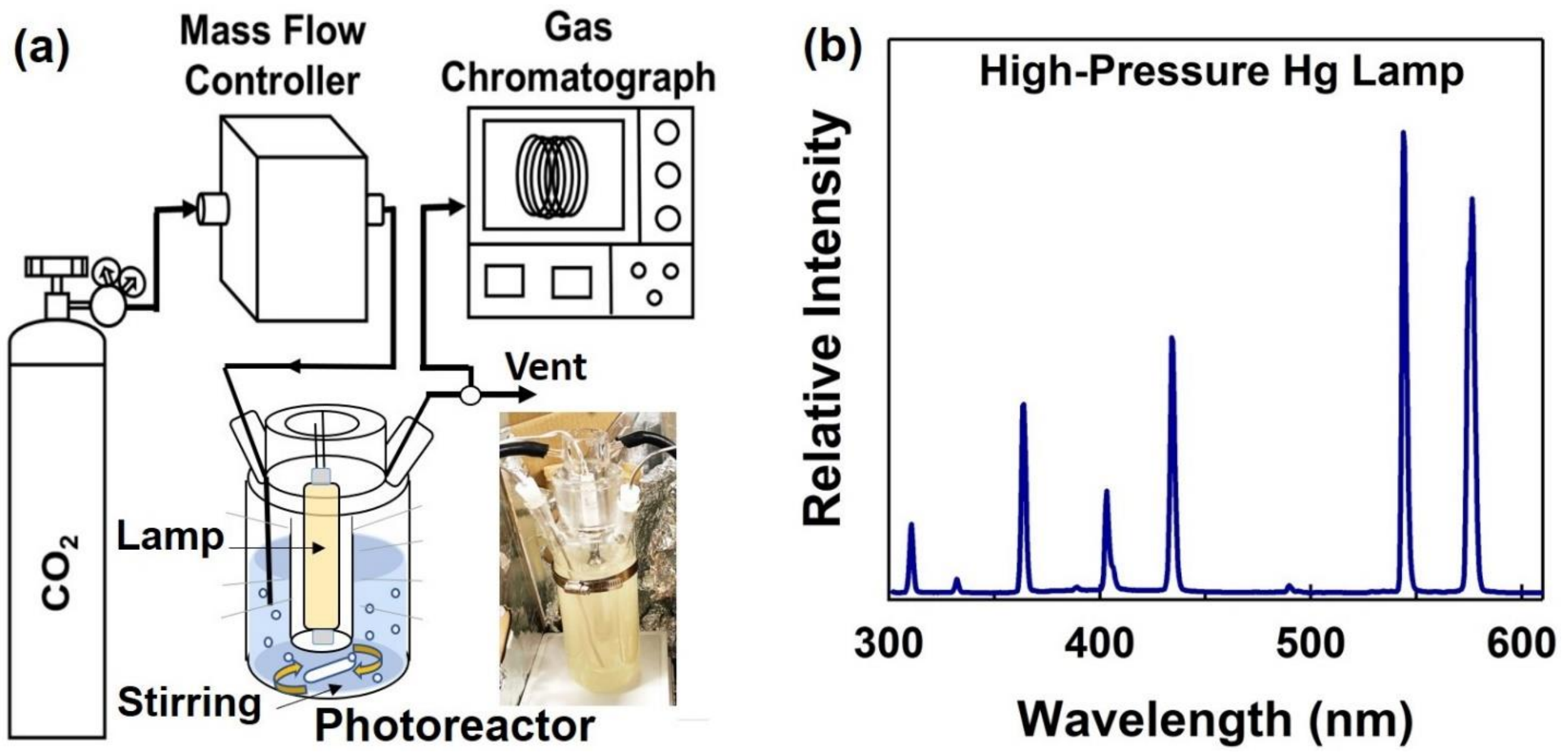

Fig. 2. (a) Description of experimental setting for photocatalytic $CO_2$ conversion including photograph of photoreactor, and (b) spectroscopy of light source used for photocatalytic test.

## 3. Results

### *3.1. Crystal structure and microstructure*

Fig. 3 shows the SEM images of HEO in various scales. Particle size measured by SEM is 25 µm. The HEO contains particles with different sizes, as shown in Fig. 3, and its specific surface area, achieved by the BET method, is 0.66 $m^2$/g. Although big size of some particles can have negative effect on photocatalytic activity due to decreasing the active surface area, this issue can be addressed in the future by using other synthesis method or advanced crushing techniques. The presence of numerous nanograins in each particle is obvious in higher magnification images in Fig. 3b, c and d. The average grain size for this material is estimated to be 192 nm, while some pores are also visible within the particles. Here, it should be noted that low specific surface area and small grain size are characteristics of materials which are synthesized/processed by the HPT method [33-36].

To confirm the successful oxidation of the material, electronic states of each element in the HEO are presented in Fig. 4 using the XPS analysis and corresponding peak deconvolution. Fig. 4 shows that the main cations in the sample are $Ti^{4+}$, $Zr^{4+}$, $Nb^{5+}$, $Hf^{4+}$ and $Ta^{5+}$, suggesting that the material is successfully oxidized to a $d^0$ electronic configuration during the high-temperature oxidation [37]. However, it should be noted that the peaks for Ti, Zr, Nb, Hf and Ta have some shoulders to the lower energy sides, suggesting that some oxygen-deficient regions with lower oxidation states should exist within the material, as confirmed by the peak deconvolution analysis (i.e., some oxygen vacancies present). The presence of vacancies is not surprising as similar issue can be observed in other HPT-processed materials due to the strain effect [35,36] and in other HEOs due to the atomic size mismatch effect [16,17].

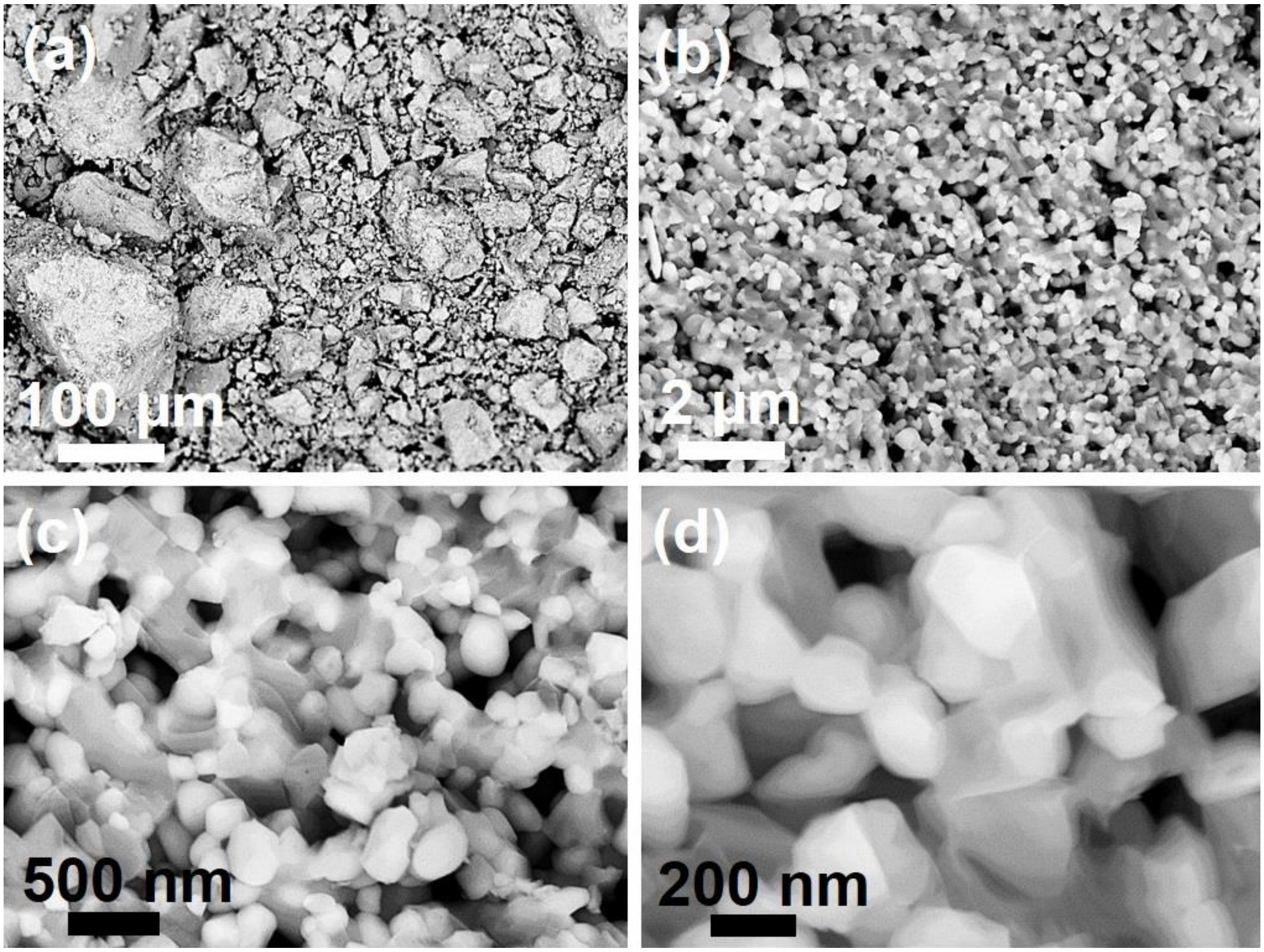

Fig. 3. Morphology of high-entropy oxide examined by SEM at different magnifications.

To confirm the distribution of elements in the material, Fig. 5a and Fig. 5b illustrate the elemental distribution mappings in the micrometer and nanometer scales, respectively. Fig. 5 shows that the elements distribute appropriately in both micrometer and nanometer scales. It is confirmed that the elements are successfully mixed by high-pressure mechanical alloying and their distribution remains reasonably homogeneous even after high-temperature oxidation. SEM-EDS analysis suggests that the material should have a general composition of $TiZrNbHfTaO_{11}$. Uniform distribution of elements is a general requirement of high-entropy materials [16-32].

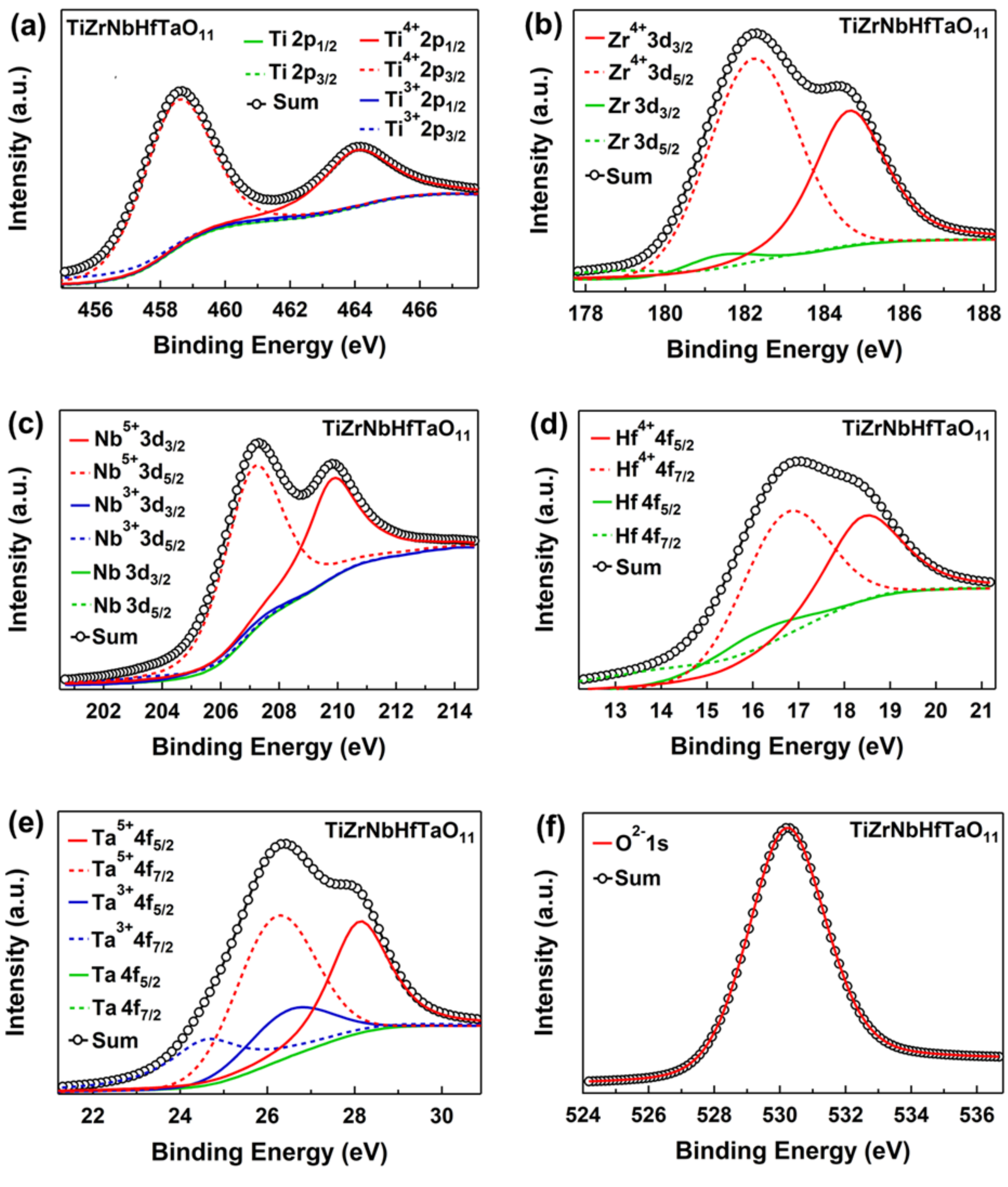

Fig. 4. Electronic states and relevant peak deconvolution of (a) Ti, (b) Zr, (c) Nb, (d) Hf, (e) Ta and (f) O in high-entropy oxide examined by XPS analysis.

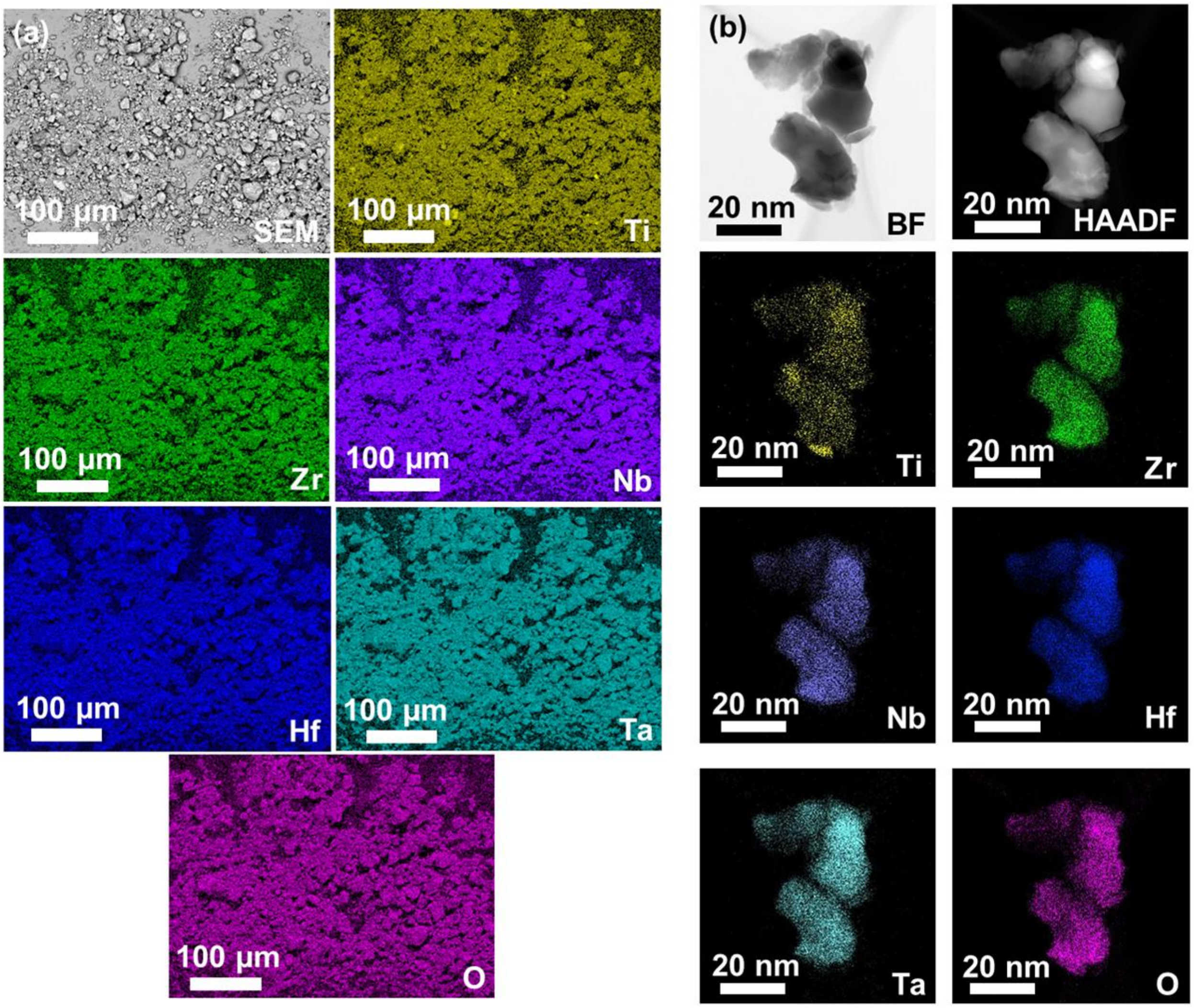

Fig. 5. Distribution of elements in high-entropy oxide examined at (a) micrometer scale using SEM-EDS and (b) nanometer scale using STEM-EDS.

Crystal structure of HEO was examined using XRD analysis, as shown in Fig. 6a. The material contains two phases with the monoclinic and orthorhombic structures. Based on the Rietveld analysis, the HEO consists of 40 wt% of monoclinic phase (A2/m space group, $a$ = 1.193 nm, $b$ = 0.381 nm, $c$ = 2.044 nm, $\alpha$ = 90°, $\beta$ = 120.16°, $\gamma$ = 90°) and 60 wt% of orthorhombic phase (Ima2 space group, $a$ = 4.092 nm, $b$ = 0.493 nm, $c$ = 0.527 nm, $\alpha = \beta = \gamma$ = 90°). Raman spectra, shown in Fig. 6b from three different positions, illustrate similar patterns in different positions, suggesting the size of phases should be smaller than the spatial resolution of micro-Raman. Taken altogether, a combination of XPS, EDS and XRD confirms that a dual-phase HEO could be successfully produced in this study.

Examination of microstructural/nanostructural features of this dual-phase HEO is shown in Fig. 7, where a is a BF image, b is a corresponding SAED pattern, c is a DF image, d and e are HR images, and f is a magnified lattice image of the selected squared region in e. Fig. 7 reveals several important points. (i) A ring pattern of SAED image confirms the presence of many nanocrystals with random orientation in Fig. 7a. (ii) The BF and DF images confirm that the grain sizes are quite small and less than 100 nm. This indicates that there are still smaller crystals within the grain-like regions observed in the SEM images of Fig. 3. (iii) The HR images confirm the co-existence of two monoclinic and orthorhombic phases at the nanoscale and large fraction of interphase boundaries. It was shown that the presence of interphases as charge heterojunctions can improve the photocatalytic activity through enhanced charge carrier separation and mobility [7]. (iv) The lattice images are quite distorted and close examination of the lattice confirms the presence of many dislocation defects within the grains. Since it was reported that the dislocations can enhance the light absorbance and photocatalytic activity at least in some semiconductors [39], the presence of dislocations in this HEO may positively act for enhancement of photocatalytic activity.

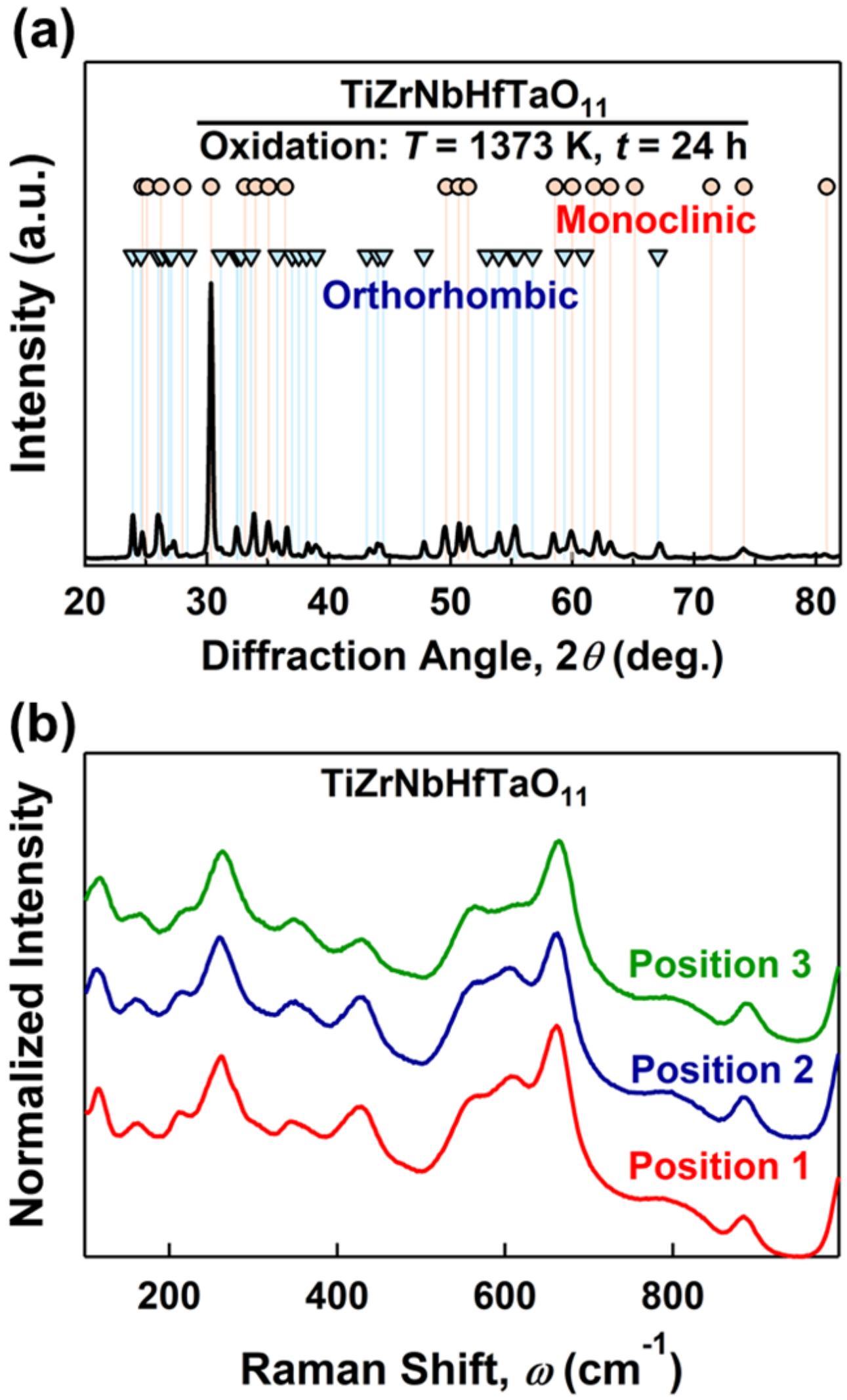

Fig. 6. Dual-phase structure of high-entropy oxide examined by (a) XRD profile and (b) micro-Raman spectra at three different positions.

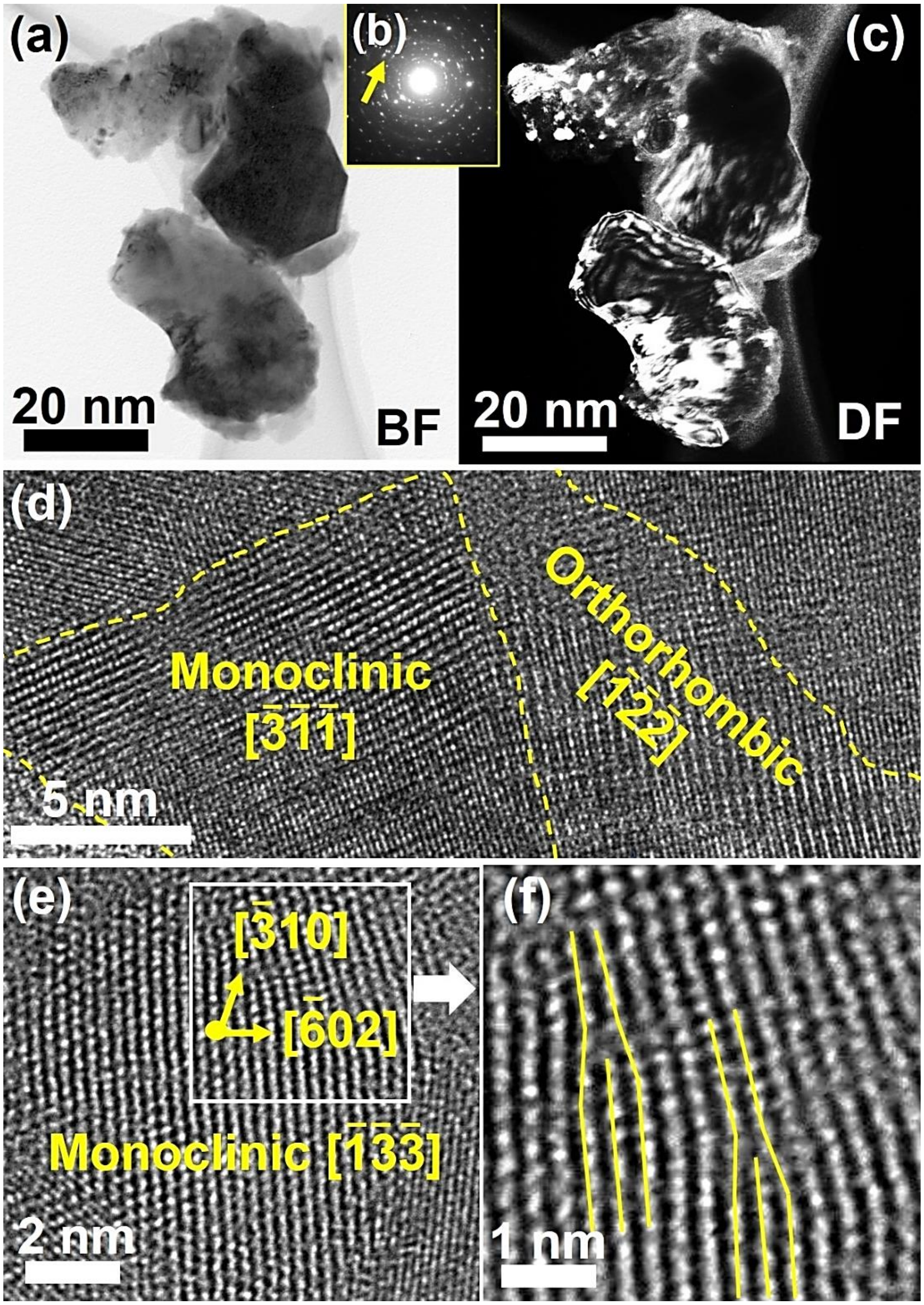

Fig. 7. Presence of nanoscaled dual phases with large fraction of interfaces and dislocations in high-entropy oxide examined by TEM (a) BF image, (b) SAED analysis, (c) DF image and (d-f) HR images, where (c) was taken with diffracted beams indicated by arrow in (a), and (f) is a magnified view of squared region in (e).

### *3.2. Electronic structure and defect states*

Fig. 8 shows (a) UV-vis absorbance spectrum, (b) Kubelka-Munk plot, (c) XPS spectrum of top of valence band and (d) electronic structure determined by a combination of UV-vis and XPS analyses. Fig. 8a indicates that the HEO can absorb light in both ultraviolet and visible light regions, although the quantity of absorbed light in the ultraviolet region is higher than that in the visible light region. Such a visible light absorbance is not detected in binary oxides such as $TiO_2$, $ZrO_2$, $HfO_2$, $Nb_2O_5$ and $Ta_2O_5$ [38,39]. Based on the Kubelka-Munk analysis, there are two apparent bandgaps of 3.0 and 2.3 eV for this HEO. The first energy gap should be related to the energy difference between the valence band and conduction band which is reasonably similar to the bandgap of $TiO_2$ and smaller than the bandgap of other binary oxides in the Ti-Zr-Hf-Nb-Ta-O system (3.1-5.7 eV) [40,41], and the second gap should be due to the defect level between the valence band and conduction band. The presence of defects (i.e., oxygen vacancies or color centers), which can be confirmed from the low energy shoulders in XPS spectra of cations, should be a main reason for the orange color of sample. The top of valence band calculated by XPS is 1.8 eV vs. NHE, which is shown by an arrow in Fig. 8c. The bottom of conduction band is calculated as -1.2 vs. NHE by considering an indirect bandgap of 3.0 eV and the defect state is estimated as -0.5 eV vs. NHE. As summarized in Fig. 8d, the potential of reactions for $CO_2$ conversion and water splitting (see Table 1) are between the energy levels for the top of valence band and the bottom of conduction band, and thus, this HEO can basically satisfy the requirements for photocatalytic reactions [5-7].

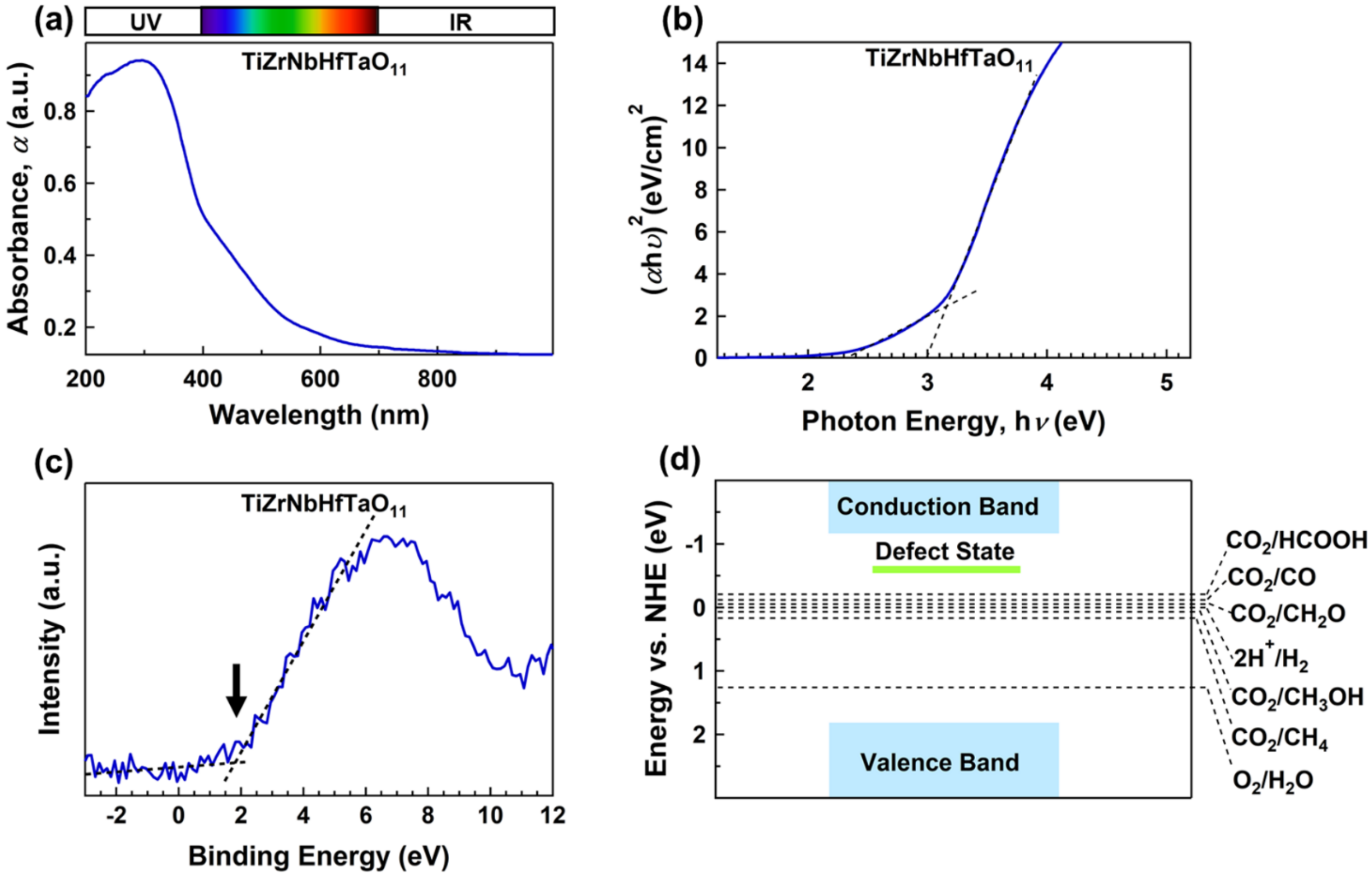

Fig. 8. Appropriate electronic structure of high-entropy oxide for photocatalytic $CO_2$ conversion. (a) UV-vis light absorbance spectrum, (b) Kubelka-Munk plot to calculate indirect bandgap ($\alpha$: light absorption, h: Planck's constant, $v$: light frequency), (c) XPS spectrum to estimate top of valence band, and (d) electronic band structure in comparison with potentials for photocatalytic $CO_2$ conversion.

### *3.3. Charge-carrier dynamics*

Charge-carrier dynamics were examined by (a) steady-state PL spectroscopy, (b) PL decay spectroscopy, (c) EPR spectroscopy and (d) photocurrent measurement, as shown in Fig. 9. The PL spectrum in Fig. 9a shows a peak at 580 nm which is equivalent to an energy level of 2.14 eV. Since this energy level is close to the energy gap of 2.3 eV, calculated using the Kubelka-Munk analysis for the defect states, it can be concluded that this peak corresponds to the recombination of excited electrons at the defect state. To have an insight into the significance of these electron-hole recombination, Table 2 compares the PL intensity and PL wavelength of the HEO with those measured by the current authors for anatase $TiO_2$ and $BiVO_4$ (as two popular photocatalysts for $CO_2$ conversion [2-5]). It is obvious that the PL intensity of HEO is lower than that of anatase $TiO_2$ and $BiVO_2$, despite its high light absorbance which is an indication of large electron-hole production. The lower PL intensity suggests that the recombination in this HEO is not higher than $TiO_2$ and $BiVO_2$, provided that the heat energy generation through the electron-hole recombination is considered identical for the three oxides.

Evaluation of PL decay intensity versus time, as shown in Fig. 9b, indicates that the PL decay of the HEO follows an exponential equation.

$$I(t) = A_1 \exp\left(-\frac{t}{\tau_1}\right) + A_2 \exp\left(-\frac{t}{\tau_2}\right) \tag{1}$$

where, $I(t)$, $A_1$, $A_2$, $\tau_1$ and $\tau_2$ are PL decay intensity at time $t$, amplitude of the first exponential function, amplitude of the second exponential function, fast decay time and slow decay time, respectively. Analysis of data in Fig. 9b suggests the values of 1.53 and 10.39 ns for $\tau_1$ and $\tau_2$, respectively. Here, the following equation can be used to estimate the average lifetime, $\tau_{ave}$ [42].

$$\tau_{ave} = \frac{A_1\tau_1^2 + A_2\tau_2^2}{A_1\tau_1 + A_1\tau_1} \tag{2}$$

Table 2 compares the average lifetime for the HEO with those for anatase $TiO_2$ and $BiVO_4$. The average lifetime for the HEO is 10.5 ns which is close to the lifetime of $TiO_2$ anatase (10.7 ns). Low recombination intensity of the HEO, measured by steady-state PL spectroscopy, and an appropriate electron lifetime close to $TiO_2$ anatase, show that the exited electrons on the surface of this material can be active for appropriate time to take part in photocatalytic reaction before recombination with holes. One reason for the appropriate charge carrier lifetime and low-intensity recombination for this HEO can be the presence of oxygen vacancies on the surface [43,44].

Prescence of oxygen vacancies, which was suggested by the orange color of sample in Fig. 1c, XPS spectroscopy in Fig. 4 and UV-vis spectroscopy in Fig. 8a, was examined further by EPR spectroscopy, as shown in Fig. 9c. Two symmetric hump peaks with a *g* factor of 2.15 appear which may be due to the oxygen vacancies, as reported in some oxides such as $Nb_2O_5$ [45]. It should be noted that the oxygen vacancies on the surface can act as active sites for electron-hole separation and photocatalytic reaction [4]. Moreover, it was shown that the surface oxygen vacancies have a direct effect on photocatalytic CO production rate: surface oxygen vacancies can absorb $CO_2$ and contribute to breaking the C=O bonds to produce CO [4].

Table 2. PL wavelength and intensity, fitted parameters of PL decay spectra and photocurrent density for high-entropy oxide in comparison with anatase $TiO_2$ and $BiVO_4$ photocatalysts.

| **PL** | | Wavelength (nm) | | Intensity (cps) | |
|---|---|---|---|---|---|
| $TiZrNbHfTaO_{11}$ | | 580 | | 190 | |
| Anatase $TiO_2$ | | 510 | | 12300 | |
| $BiVO_4$ | | 640 | | 300 | |
| **PL Decay** | $\tau_1$ (ns) | $\tau_2$ (ns) | $A_1$ | $A_2$ | $\tau_{ave}$ (ns) |
| $TiZrNbHfTaO_{11}$ | 1.53 | 10.39 | 42.34 | 57.66 | 10.5 |
| Anatase $TiO_2$ | 1.24 | 11.46 | 41.98 | 58.02 | 10.7 |
| $BiVO_4$ | 2.17 | 14.90 | 56.56 | 43.44 | 12.9 |
| **Photocurrent ($mA/m^2$)** | | Cycle 1 | Cycle 2 | Cycle 3 | Cycle 4 |
| $TiZrNbHfTaO_{11}$ | | 9.6 | 8.9 | 8.4 | 8.2 |
| Anatase $TiO_2$ | | 43.5 | 32.2 | 27.9 | 25.2 |
| $BiVO_4$ | | 18.6 | 17.1 | 17.0 | 16.7 |

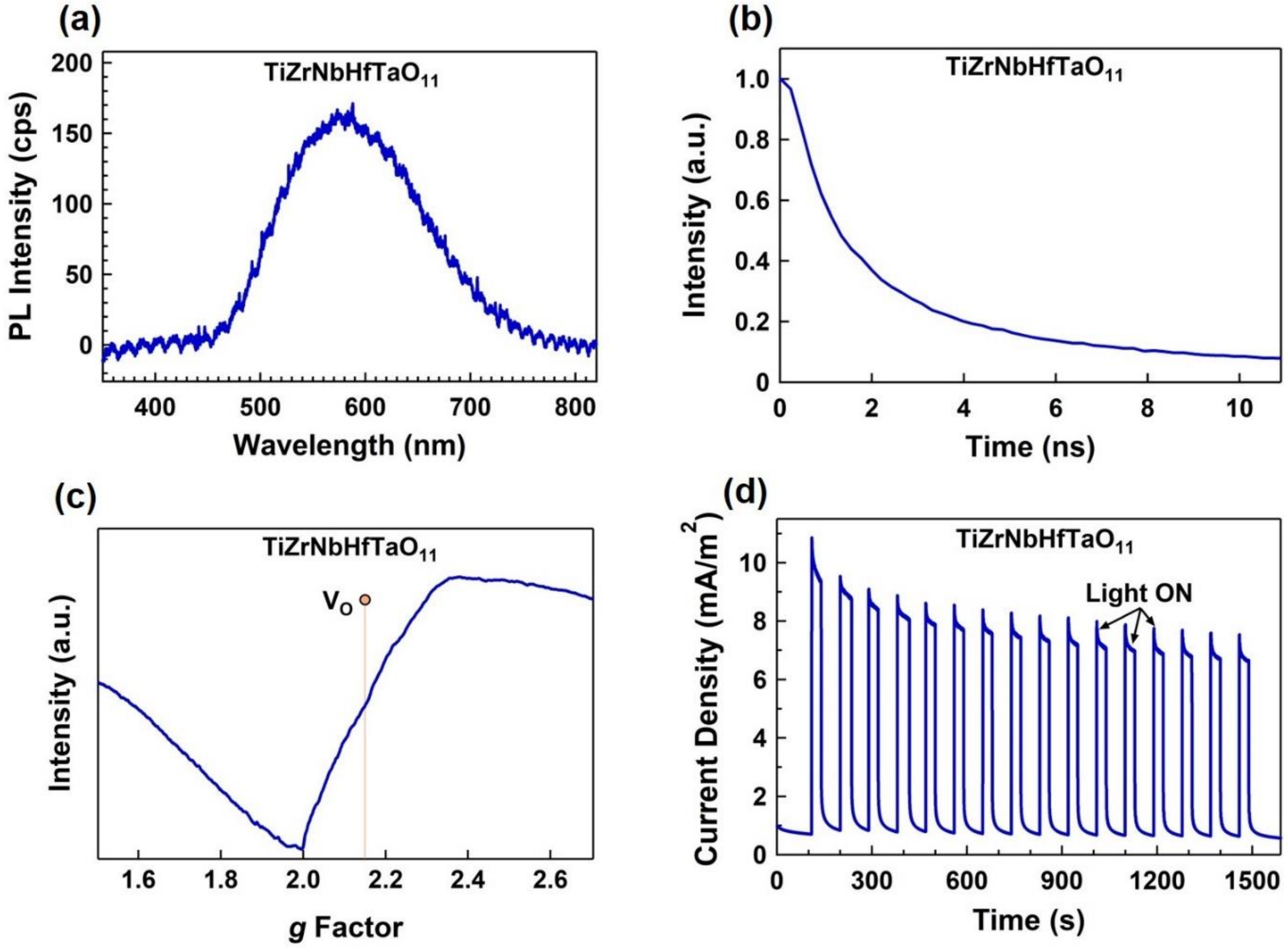

Fig. 9. (a) Steady-state PL emission, (b) time-resolved PL decay, (c) EPR spectra and (d) photocurrent generation for high-entropy oxide.

Fig. 9d shows photocurrent measurement on HEO thin film. The material successfully generates photocurrent, although its photocurrent density decreases during the time due to the accumulation of holes with positive charge on the surface. Table 2 compares the photocurrent density of the HEO with that of reference anatase $TiO_2$ and $BiVO_4$ for the first four cycles. It should be noted that the quantitative comparison of the photocurrent density of these three materials should be conducted by care due to the technical limits in making dense films with good FTO-oxide bonding by annealing at 473 K. The photocurrent density of HEO is apparently lower than that of the reference oxides. Despite the low photocurrent density of HEO, photocurrent generation on this material indicates that the exited electrons can have enough lifetime to separate from the surface of material and take part in the photocurrent generation. The generation of photocurrent is a positive sign for possible photocatalytic activity of this HEO, as discussed earlier for other photocatalysts [38].

***3.4. Photocatalytic activity***

Photocatalytic activity of HEO for $CO_2$ conversion is summarized in Fig. 10. As shown in Fig. 10a and b, the HEO could successfully produce both CO and $H_2$ under the full arc emission of high-pressure Hg lamp without any co-catalyst addition, despite its low specific surface area as 0.66 $m^2/g$ (the error bar of gas amount measurement for three repeated tests was lower than 10%). Independent synthesis of the HEO material and repeating the photocatalytic test, as indicated as Sample #2 in Fig. 10a, also confirm the high activity of this material for photocatalytic $CO_2$ conversion with a reasonably constant CO and $H_2$ production rate within an extended irradiation time of 10 h. The amount of CO production is higher and the amount of $H_2$ production is lower for Sample #2 compared to Sample #1, suggesting that the activity of this HEO can be still improved by modification of the synthesis method. Two points should be noted here. First, CO and $H_2$ were the only reaction products within the detection limits of analyses and no other products including methane could be detected. Second, blank tests confirmed that no CO and $H_2$ are produced by (i) $CO_2$ injection in the presence of HEO under the dark condition for 2 h, (ii) Ar injection in the presence of HEO under the light irradiation for 1 h, and (iii) $CO_2$ injection without the presence of HEO under the light irradiation for 5 h. The stability of HEO, examined by XRD analysis after the photocatalytic test, is shown in Fig. 10c, indicating that the crystal structure of the HEO is stable after photocatalytic test. The stability of HEOs, which was also reported for other applications such thermal barrier coatings [20,21], magnetic components [22,23], dielectric components [24,25], Li-ion batteries [26,27], Li-S batteries [28], Zn-air batteries [29], catalysts [30,31] and electrocatalysts [32], is usually due to their low Gibbs free energy resulting from their high entropy [16,17].

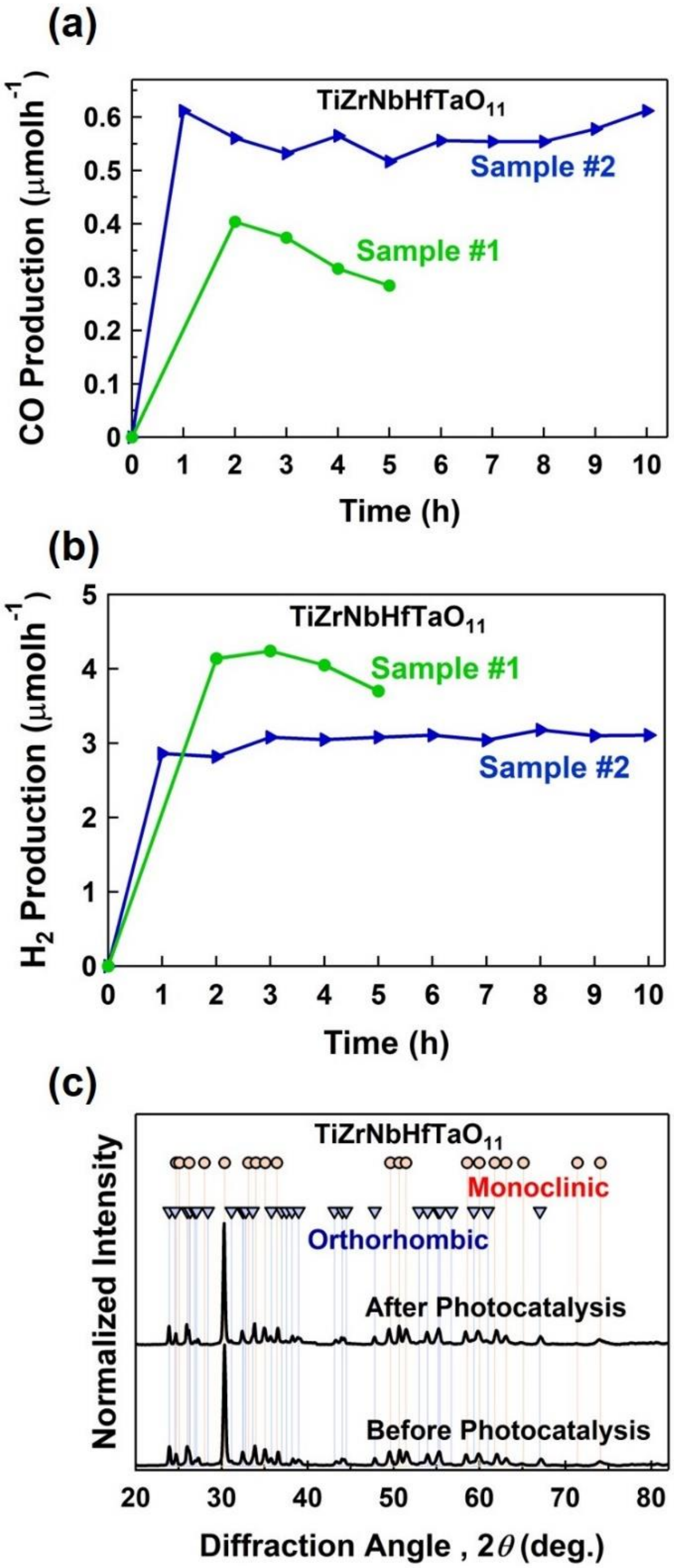

Fig. 10. Photocatalytic activity of high-entropy oxide for $CO_2$ conversion and $H_2O$ decomposition. (a) CO production rate versus time, (b) $H_2$ production rate versus time, and (c) XRD pattern before and after photocatalytic test.

## 4. Discussion

Three issues need to be discussed in detail here: (i) comparison of photocatalytic activity of the HEO with available photocatalysts, (ii) factors influencing the photocatalytic activity of the HEO, and (iii) mechanism of $CO_2$ conversion on the HEO photocatalyst.

Although the current results confirm the potential of HEOs as a new family of photocatalysts for $CO_2$ conversion, their activity should be compared with other photocatalysts to have an insight into their significance. To understand this issue, photocatalytic $CO_2$ conversion activity of the HEO with a specific surface area of 0.66 $m^2/g$ was compared with anatase $TiO_2$ (99.8%), $BiVO_4$ (99.9%) and P25 $TiO_2$ (99.5%) with the surface areas of 10.2, 0.3 and 38.7 $m^2/g$, respectively. Since various parameter such as catalyst concentration, temperature, reactor type, light source type and light intensity can influence the CO production rate, photocatalytic activity of these materials were compared in the same conditions. Fig. 11 shows the activity of these materials per 1 g of catalyst. The CO production rate for HEO is significantly higher than anatase $TiO_2$ and $BiVO_4$ which are some of the most popular photocatalysts for photocatalytic $CO_2$ conversion. Moreover, the CO and $H_2$ production rate on this HEO is comparable with P25 $TiO_2$ as a benchmark photocatalyst, although the surface area of current HEO is 60 times smaller than that of P25 $TiO_2$. It should be noted that the quantity of $H_2$ production on anatase $TiO_2$ and $BiVO_4$ was not within the detection limits of gas chromatograph. To get more insight on the significance of photocatalytic $CO_2$ conversion on this HEO, its activity was compared with some reported data in the literature [8,9,11,46-53]. Although the experiments in the literature are not conducted under a consistent and standard condition, it is still useful to have a comparison. As given in Table 3, the amount of CO production rate varies in a wide range of 0.12-10.16 $\mu molh^{-1}$ $g^{-1}$. The average CO production for HEO is 4.64±0.30 $\mu molh^{-1}g^{-1}$ which is higher than many of the reported values in Table 3.

The reason for high CO production rate on current HEO can be attributed to various factors: the presence of lattice defects such as oxygen vacancies which can act as activation sites [9,12], the presence of five cations which can enhance the activity by straining effect [14,15], appropriate electronic structure which satisfy most of the reactions for $CO_2$ conversion and water splitting [5,6], and appropriate lifetime of charge carriers to participate in photocatalytic reaction due to the defective nature of HEOs [43,44]. Moreover, the presence of two phases can improve the charge carrier separation through interfaces and enhance the photocatalytic activity [7,38]. The presence of several cations in the HEO can also produce hybridized orbitals with higher activity for chemical reactions [16,17]. To further enhance the efficiency of current HEO for photocatalytic $CO_2$ conversion, future works are required to enhance its specific surface area by improving the synthesis or crushing techniques.

Regarding the third issue, three main mechanisms for photocatalytic $CO_2$ reduction have been suggested, as summarized in Table 4: carbene pathway, formaldehyde pathway and glyoxal pathway [3,4]. The behavior of current HEO is similar to P25 $TiO_2$, suggesting that both materials probably follow the same pathway. Although even for $TiO_2$ with different impurities and lattice defects, there are still significant arguments regarding the $CO_2$ reduction pathways, it is still possible to discuss about the possible mechanisms for current HEO photocatalyst. The nonappearance of HCOOH, $CH_3OH$ and $CH_4$ in the gas and liquid phases within the detection limits of analyses suggests that the formaldehyde pathway may not be the major mechanism [3,4]. The nonappearance of HCOOH and $CH_4$ also indicates that the glyoxal pathway may not be the major mechanism [3,4]. The production of CO suggests that the carbene pathway is probably the major mechanism. However, the absence of $CH_4$ and the presence of $H_2$, which is similar to the behavior of P25 $TiO_2$ in this study, indicates that the carbene pathway possibly stops at some intermediate stages due to the formation of $H_2$ gas [54]. The absence of $CH_4$ can also be explained by the defective structure of HEO. Since the HEO material has oxygen vacancies as surface defects, $CO_2$ in connection with $H_2O$ as a Lewis acid tends to adsorb on oxygen vacancies [54].

This adsorption degrades C=O bonding and produce •CO radicals and consequently generates CO gas [4]. Compared with $CO_2$, the generated CO has lower tendency to be adsorbed on the surface defects [54], and thus, the carbene pathway does no continue to produce detectable quantity of $CH_4$. For $TiO_2$, it was also reported that although the $CH_4$ formation in the carbene pathway is thermodynamically more favorable than CO and $H_2$ formation, the formation of $CH_4$ is kinetically more difficult because it needs higher numbers of electrons and protons [55].

Taken altogether, this study introduces HEOs as active photocatalysts for $CO_2$ conversion, and this opens a path to explore numerous photocatalysts by considering the state-of-art on engineering of catalysts for $CO_2$ photoreduction [56]. Despite high activity of current HEO, future studies are required to clarify the exact $CO_2$ conversion mechanism on this new family of materials. It should be noted that although the material in this study was synthesized by a two-step high-pressure mechanical alloying and high-temperature oxidation, other methods developed earlier for the synthesis of high-entropy ceramics [57] can be used in the future to synthesize powders with high specific surface area and low economical cost. Moreover, since earlier studies showed that the semiconductor photocatalysts with $CO_2$ conversion capability can have good activity for degradation of organic pollutants as well [58,59], it is expected that the photocatalytic activity of HEOs is not limited to CO and $H_2$ production.

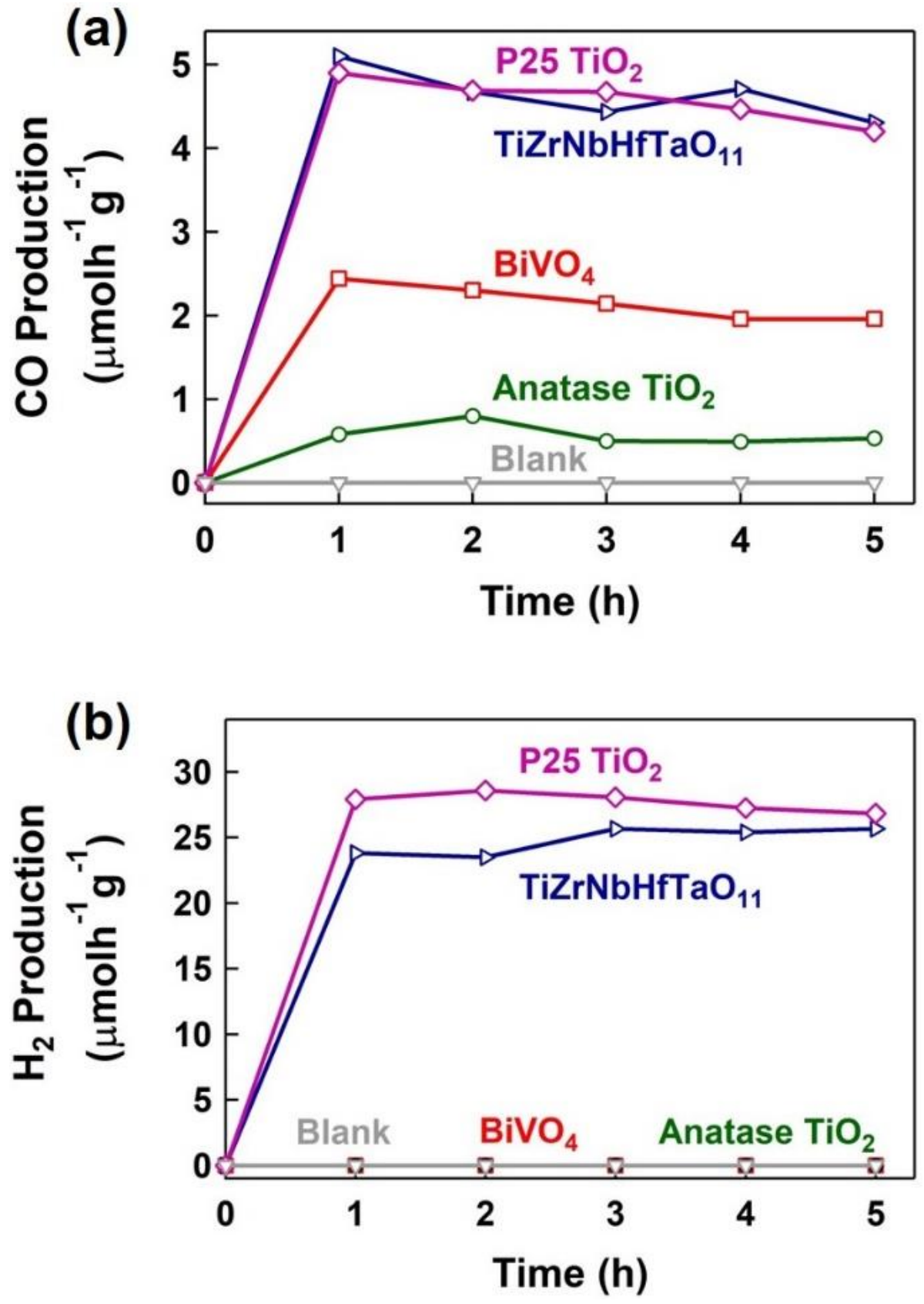

Fig. 11. High efficiency of high-entropy oxide compared with $TiO_2$ and $BiVO_4$ for photocatalytic $CO_2$ conversion. (a) CO production rate and (b) $H_2$ production rate versus time.

Table 3. Summary of some reported photocatalytic $CO_2$ conversion rates in literature in comparison with results of current study.

| Photocatalyst | Light Source | CO Production ($\mu molh^{-1}g^{-1}$) | References |
|---|---|---|---|
| $TiO_2$ Nanosheet -CN | 150 W Xe lamp | 2.04 | [8] |
| $TiO_2$- Graphitic carbon | 300 W Xe lamp | 10.16 | [46] |
| $TiO_2$ nanosheets exposed {001} facet | 2 *18W Hg lamps | 0.12 | [47] |
| $TiO_2$ - Hydrogenated $CoO_x$ | 150W UV lamp | 1.24 | [48] |
| $TiO_2$ 3D Ordered Microporous - Pd | 300 W Xe lamp | 3.9 | [49] |
| $C_3N_4$ by Thermal Condensation | 350 W Xe lamp | 4.83 | [9] |
| $Cd_{1-x}Zn_xS$ | 100 W LED plate | 2.9 | [50] |
| BiOI | 300 W Xe lamp | 4.1 | [51] |
| $xCu_2O$-$Zn_{2-2x}Cr$ | 200-W Hg-Xe lamp | 2.5 | [11] |
| $CeO_{2-x}$ | 300 W Xe lamp | 1.65 | [52] |
| $Cu_2O/RuO_x$ | 150 W Xe lamp | 0.88 | [53] |
| $TiO_2$ Anatase | 400 W Hg Lamp | 0.58±0.12 | This Work |
| $BiVO_4$ | 400 W Hg Lamp | 2.16±0.21 | This Work |
| $TiO_2$ P25 | 400 W Hg Lamp | 4.63±0.33 | This Work |
| $TiZrNbHfTaO_{11}$ | 400 W Hg Lamp | 4.64±0.30 | This Work |

Table 4. Main mechanisms for $CO_2$ photocatalytic reduction pathway [3].

| Carbene Pathway | Formaldehyde Pathway | Glyoxal Pathway |
|---|---|---|
| (1) $CO_2 + e^- \rightarrow CO_2^{\bullet -}$ | (1) $CO_2 + e^- \rightarrow CO_2^{\bullet -}$ | (1) $CO_2 + e^- \rightarrow CO_2^{\bullet -}$ |
| (2) $CO_2^{\bullet -} + e^- + H^+ \rightarrow CO + OH^-$ | (2) $CO_2^{\bullet -} + H^+ \rightarrow {}^{\bullet}COOH$ | (2) $CO_2^{\bullet -} + e^- + H^+ \rightarrow CHOO^-$ |
| (3) $CO + e^- \rightarrow CO^{\bullet -}$ | (3) ${}^{\bullet}COOH + e^- + H^+ \rightarrow HCOOH$ | (3) $CHOO^- + H^+ \rightarrow HCOOH$ |
| (4) $CO^{\bullet -} + e^- + H^+ \rightarrow C + OH^-$ | (4) $HCOOH + e^- + H^+ \rightarrow H_3OOC^{\bullet}$ | (4) $HCOOH + e^- \rightarrow HOC^{\bullet}$ |
| (5) $C + e^- + H^+ \rightarrow CH^{\bullet}$ | (5) $HCOOH_2^{\bullet} + e^- + H^+ \rightarrow HCOH + H_2O$ | (5) $HOC^{\bullet} + OH^- \rightarrow C_2H_2O_2$ |
| (6) $CH^{\bullet} + e^- + H^+ \rightarrow CH_2$ | (6) $HCOH + e^- \rightarrow H_2C^{\bullet}O^-$ | (6) $C_2H_2O_2 + e^- + H^+ \rightarrow H_3O_2C_2^{\bullet}$ |
| (7) $CH_2 + e^- + H^+ \rightarrow CH_3^{\bullet}$ | (7) $H_2C^{\bullet}O^- + H^+ \rightarrow H_2OHC^{\bullet}$ | (7) $H_3O_2C_2^{\bullet} + e^- + H^+ \rightarrow C_2H_4O_2$ |
| (8) $CH_3^{\bullet} + e^- + H^+ \rightarrow CH_4$ | (8) $H_2OHC^{\bullet} + e^- + H^+ \rightarrow CH_3OH$ | (8) $C_2H_4O_2 + e^- + H^+ \rightarrow H_3OC_2^{\bullet} + H_2O$ |
| (9) $CH_3^{\bullet} + OH^- \rightarrow CH_3OH$ | (9) $CH_3OH + e^- + H^+ \rightarrow {}^{\bullet}CH_3 + H_2O$ | (9) $H_3OC_2^{\bullet} + e^- + H^+ \rightarrow C_2H_4O$ |
| | (10) ${}^{\bullet}CH_3 + e^- + H^+ \rightarrow CH_4$ | (10) $C_2H_4O + h^+ \rightarrow H_3OC_2^{\bullet} + H^+$ |
| | | (11) $H_3OC_2^{\bullet} \rightarrow CH_3^{\bullet} + CO$ |
| | | (12) $CH_3^{\bullet} + e^- + H^+ \rightarrow CH_4$ |

## 5. Conclusion

A high-entropy oxide with a general composition of $TiZrNbHfTaO_{11}$ was synthesized and used for photocatalytic $CO_2$ conversion. Due to appropriate electronic band structure, good charge carrier lifetime and a defective and strained dual-phase structure, the material acted as a photocatalyst for $CO_2$ to CO conversion and $H_2O$ to $H_2$ production without addition of any co-catalyst. The photocatalytic activity of this oxide was better than well-known anatase $TiO_2$ and $BiVO_4$ photocatalysts and comparable with P25 $TiO_2$ as a benchmark photocatalyst, suggesting high-entropy oxides as a new family of photocatalysts for $CO_2$ conversion.

**Acknowledgments**

This work is supported in part by the WPI-I2CNER, Japan, and in part by Grants-in-Aid for Scientific Research on Innovative Areas from the MEXT, Japan (19H05176 & 21H00150).

**References**

[1] M. Forkel, N. Carvalhais, C. Rödenbeck, R. Keeling, M. Heimann, K. Thonicke, S. Zaehle, M. Reichstein, Enhanced seasonal $CO_2$ exchange caused by amplified plant productivity in northern ecosystems, Science 351 (2016) 696-699. https://doi.org/10.1126/science.aac4971.

[2] A.J. Morris, G.J. Meyer, E. Fujita, Molecular approaches to the photocatalytic reduction of carbon dioxide for solar fuels, Acc. Chem. Res. 42 (2009) 1983-1994. https://doi.org/10.1021/ar9001679.

[3] S.N. Habisreutinger, L. Schmidt-Mende, J.K. Stolarczyk, Photocatalytic reduction of $CO_2$ on $TiO_2$ and other semiconductors, Angew. Chem. Int. Ed. 52 (2013) 7372-7408. https://doi.org/10.1002/anie.201207199.

[4] K. Wang, J. Lu, Y. Lu, C.H. Lau, Y. Zheng, X. Fan, Unravelling the C-C coupling in $CO_2$ photocatalytic reduction with $H_2O$ on Au/$TiO_{2-x}$: combination of plasmonic excitation and oxygen vacancy, Appl. Catal. B 292 (2021) 120147. https://doi.org/10.1016/j.apcatb.2021.120147.

[5] X. Li, J. Yu, M. Jaroniec, Hierarchical photocatalysts, Chem. Soc. Rev. 45 (2016) 2603-2636. https://doi.org/10.1039/C5CS00838G.

[6] E. Kalamaras, M.M. Maroto-Valer, M. Shao, J. Xuan, H. Wang, Solar carbon fuel via photoelectrochemistry, Catal. Today 317 (2018) 56-75. https://doi.org/10.1016/j.cattod.2018.02.045.

[7] K. Li, B. Peng, T. Peng, Recent advances in heterogeneous photocatalytic $CO_2$ conversion to solar fuels, ACS Catal. 6 (2016) 7485-7527. https://doi.org/10.1021/acscatal.6b02089.

[8] A. Crake, K.C. Christoforidis, R. Godin, B. Moss, A. Kafizas, S. Zafeiratos, J.R. Durrant, C. Petit, Titanium dioxide/carbon nitride nanosheet nanocomposites for gas phase $CO_2$ photoreduction under UV-visible irradiation, Appl. Catal. B 242 (2019) 369-378. https://doi.org/10.1016/j.apcatb.2018.10.023.

[9] P. Xia, M. Antonietti, B. Zhu, T. Heil, J. Yu, S. Cao, Designing defective crystalline carbon nitride to enable selective $CO_2$ photoreduction in the gas phase, Adv. Funct. Mater. 29 (2019) 1900093. https://doi.org/10.1002/adfm.201900093.

[10] J. Bian, J. Feng, Z. Zhang, Z. Li, Y. Zhang, Y. Liu, S. Ali, Y. Qu, L. Bai, J. Xie, D. Tang, X. Li, F. Bai, J. Tang, L. Jing, Dimension-matched zinc phthalocyanine/$BiVO_4$ ultrathin nanocomposites for $CO_2$ reduction as efficient wide-visible-light-driven photocatalysts via a cascade charge transfer, Angew. Chem. 131 (2019) 10989-10994. https://doi.org/10.1002/ange.201905274.

[11] H. Jiang, K. Katsumata, J. Hong, A. Yamaguchi, K. Nakata, C. Terashima, N. Matsushita, M. Miyauchi, A. Fujishima, Photocatalytic reduction of $CO_2$ on $Cu_2O$-loaded Zn-Cr layered double hydroxides, Appl. Catal. B 224 (2018) 783-790. https://doi.org/10.1016/j.apcatb.2017.11.011.

[12] J.J. Li, M. Zhang, B. Weng, X. Chen, J. Chen, H.P. Jia, Oxygen vacancies mediated charge separation and collection in Pt/$WO_3$ nanosheets for enhanced photocatalytic performance, Appl. Surf. Sci. 507 (2020) 145133. https://doi.org/10.1016/j.apsusc.2019.145133.

[13] Y. Li, W.N. Wang, Z. Zhan, M.H. Woo, C.Y. Wu, P. Biswas, Photocatalytic reduction of $CO_2$ with $H_2O$ on mesoporous silica supported $Cu/TiO_2$ catalysts, Appl. Catal. B 100 (2010) 386-392. https://doi.org/10.1016/j.apcatb.2010.08.015.
[14] Z. Liu, C. Menéndez, J. Shenoy, J.N. Hart, C.C. Sorrell, C. Cazorl, Strain engineering of oxide thin films for photocatalytic applications, Nano Energy 72 (2020) 104732. https://doi.org/10.1016/j.nanoen.2020.104732.
[15] J. Di, P. Song, C. Zhu, C. Chen, J. Xiong, M. Duan, R. Long, W. Zhou, M. Xu, L. Kang, B. Lin, D. Liu, S. Chen, C. Liu, H. Li, Y. Zhao, S. Li, Q. Yan, L. Song, Z. Liu, Strain-engineering of $Bi_2O_{17}Br_2$ nanotubes for boosting photocatalytic $CO_2$ reduction, ACS Mater. Lett. 2 (2020)1025–1032. https://doi.org/10.1021/acsmaterialslett.0c00306.
[16] C. Oses, C. Toher, S. Curtarolo, High-entropy ceramics, Nat. Rev. Mater. 5 (2020) 295-309. https://doi.org/10.1038/s41578-019-0170-8.
[17] A.J. Wright, Q. Wang, C. Huang, A. Nieto, R. Chen, J. Luo, From high-entropy ceramics to compositionally-complex ceramics: a case study of fluorite oxides, J. Eur. Ceram. Soc. 40 (2020) 2120-2129. https://doi.org/10.1016/j.jeurceramsoc.2020.01.015.
[18] H. Chen, K. Jie, C.J. Jafta, Z. Yang, S. Yao, M. Liu, Z. Zhang, J. Liu, M. Chi, J. Fu, S. Dai, An ultrastable heterostructured oxide catalyst based on high-entropy materials: a new strategy toward catalyst stabilization via synergistic interfacial interaction, Appl. Catal. B 276 (2020) 119155. https://doi.org/10.1016/j.apcatb.2020.119155.
[19] H. Chen, J. Fu, P. Zhang, H. Peng, C.W. Abney, K. Jie, X. Liu, M. Chi, S. Dai, Entropy-stabilized metal oxide solid solutions as CO oxidation catalysts with high-temperature stability, J. Mater. Chem. A 6 (2018) 11129-11133. https://doi.org/10.1039/C8TA01772G.
[20] J.L. Braun, C.M. Rost, M. Lim, A. Giri, D.H. Olson, G.N. Kotsonis, G. Stan, D.W. Brenner, J.P. Maria, P.E. Hopkins, Charge-induced disorder controls the thermal conductivity of entropy-stabilized oxides, Adv. Mater. 30 (2018) 1805004. https://doi.org/10.1002/adma.201805004.
[21] A.J. Wright, C. Huang, M.J. Walock, A. Ghoshal, M. Murugan, J. Luo, Sand corrosion, thermal expansion, and ablation of medium-and high-entropy compositionally complex fluorite oxides, J. Am. Ceram. Soc. 104 (2021) 448-462. https://doi.org/10.1111/jace.17448.
[22] R. Witte, A. Sarkar, R. Kruk, B. Eggert, R.A. Brand, H. Wende, H. Hahn, High-entropy oxides: an emerging prospect for magnetic rare-earth transition metal perovskites, Phys. Rev. Mater. 3 (2019) 034406. https://doi.org/10.1103/PhysRevMaterials.3.034406.
[23] A. Mao, H.Z. Xiang, Z.G. Zhang, K. Kuramoto, H. Zhang, Y. Jia, A new class of spinel high-entropy oxides with controllable magnetic properties, J. Magn. Magn. Mater. 497 (2020) 165884. https://doi.org/10.1016/j.jmmm.2019.165884.
[24] A. Radoń, Ł. Hawełek, D. Łukowiec, J. Kubacki, P. Włodarczyk, Dielectric and electromagnetic interference shielding properties of high entropy $(Zn,Fe,Ni,Mg,Cd)Fe_2O_4$ ferrite, Sci. Rep. 9 (2019) 20078. https://doi.org/10.1038/s41598-019-56586-6.
[25] S. Zhou, Y. Pu, Q. Zhang, R. Shi, X. Guo, W. Wang, J. Ji, T. Wei, T. Ouyang, Microstructure and dielectric properties of high entropy $Ba(Zr_{0.2}Ti_{0.2}Sn_{0.2}Hf_{0.2}Me_{0.2})O_3$ perovskite oxides, Ceram. Int. 46 (2020) 7430-7437. https://doi.org/10.1016/j.ceramint.2019.11.239.
[26] A. Sarkar, L. Velasco, D. Wang, Q. Wang, G. Talasila, L. de Biasi, C. Kübel, T. Brezesinski, S.S. Bhattacharya, H. Hahn, B. Breitung, High entropy oxides for reversible energy storage, Nat. Commun. 9 (2018) 3400. https://doi.org/10.1038/s41467-018-05774-5.

[27] T.X. Nguyen, J. Patra, J.K. Chang, J.M. Ting, High entropy spinel oxide nanoparticles for superior lithiation-delithiation performance, J. Mater. Chem. A 8 (2020) 18963-18973. https://doi.org/10.1039/D0TA04844E.
[28] Y. Zheng, Y. Yi, M. Fan, H. Liu, X. Li, R. Zhang, M. Li, Z.-A. Qiao, A high-entropy metal oxide as chemical anchor of polysulfide for lithium-sulfur batteries, Energy Storage Mater. 23 (2019) 678-683. https://doi.org/10.1016/j.ensm.2019.02.030.
[29] G. Fang, J. Gao, J. Lv, H. Jia, H. Li, W. Liu, G. Xie, Z. Chen, Y. Huang, Q. Yuan,X. Liu, X. Lin, S. Sun, H.J. Qiu, Multi-component nanoporous alloy/(oxy) hydroxide for bifunctional oxygen electrocatalysis and rechargeable Zn-air batteries, Appl. Catal. B 268 (2020) 118431. https://doi.org/10.1016/j.apcatb.2019.118431.
[30] H. Chen, W. Lin, Z. Zhang, K. Jie, D.R. Mullins, X. Sang, S.-Z. Yang, C.J. Jafta, C.A. Bridges, X. Hu, Mechanochemical synthesis of high entropy oxide materials under ambient conditions: dispersion of catalysts via entropy maximization, ACS Mater. Lett. 1 (2019) 83-88. https://doi.org/10.1021/acsmaterialslett.9b00064.
[31] M.S. Lal, R. Sundara, High entropy oxides - a cost-effective catalyst for the growth of high yield carbon nanotubes and their energy applications, ACS Appl. Mater. Interfaces 11 (2019) 30846-30857. https://doi.org/10.1021/acsami.9b08794.
[32] T.X. Nguyen, Y.C. Liao, C.C. Lin, Y.H. Su, J.M. Ting, Advanced high entropy perovskite oxide electrocatalyst for oxygen evolution reaction, Adv. Funct. Mater. 31 (2021) 2101632. https://doi.org/10.1002/adfm.202101632
[33] P. Edalati, Q. Wang, H. Razavi-Khosroshahi, M. Fuji, T. Ishihara, K. Edalati, Photocatalytic hydrogen evolution on a high-entropy oxide, J. Mater. Chem. A 8 (2020) 3814-3821. https://doi.org/10.1039/C9TA12846H.
[34] P. Edalati, X.F. Shen, M. Watanabe, T. Ishihara, M. Arita, M. Fuji, K. Edalati, High-entropy oxynitride as a low-bandgap and stable photocatalyst for hydrogen production, J. Mater. Chem. A 9 (2021) 15076-15086. https://doi.org/10.1039/D1TA03861C.
[35] K. Edalati, Z. Horita, A review on high-pressure torsion (HPT) from 1935 to 1988, Mater. Sci. Eng. A 652 (2016) 325–352. https://doi.org/10.1016/j.msea.2015.11.074.
[36] K. Edalati, Review on recent advancements in severe plastic deformation of oxides by high-pressure torsion (HPT), Adv. Eng. Mater. 21 (2019) 1800272. https://doi.org/10.1002/adem.201800272.
[37] J. Chastain, Handbook of X-ray Photoelectron Spectroscopy, Perkin-Elmer Corporation, Eden Prairie, MN, USA, 1992.
[38] S. Akrami, M. Watanabe, T.H. Ling, T. Ishihara, M. Arita, M. Fuji, K. Edalati, High-pressure $TiO_2$-II polymorph as an active photocatalyst for $CO_2$ to CO conversion, Appl. Catal. B 298 (2021) 120566. https://doi.org/10.1016/j.apcatb.2021.120566.
[39] L. Ran, J. Hou, S. Cao, Z. Li, Y. Zhang, Y. Wu, B. Zhang, P. Zhai, L. Sun, Defect engineering of photocatalysts for solar energy conversion, Sol. RRL 4 (2020) 1900487. https://doi.org/10.1002/solr.201900487.
[40] M. R. Hoffmann, S. T. Martin, W. Choi and D. W. Bahnemann, Environmental applications of semiconductor photocatalysis, Chem. Rev. 95 (1995) 69–96. https://doi.org/10.1021/cr00033a004.
[41] K. Maeda and K. Domen, New non-oxide photocatalysts designed for overall water splitting under visible light, J. Phys. Chem. C, 111 (2007) 7851–7861. https://doi.org/10.1021/jp070911w.

[42] Z. Zhang, K. Liu, Z. Feng, Y. Bao, B. Dong, Hierarchical sheet-on-sheet $ZnIn_2S_4/g-C_3N_4$ heterostructure with highly efficient photocatalytic $H_2$ production based on photoinduced interfacial charge transfer, Sci. Rep. 6 (2016) 19221. https://doi.org/10.1038/srep1922.
[43] X.H. Wang, J.G. Li, H. Kamiyama, M. Katada, N. Ohashi, Y. Moriyoshi, T. J. Ishigaki, Pyrogenic iron (III)-doped $TiO_2$ nanopowders synthesized in RF thermal plasma: phase formation, defect structure, band gap, and magnetic properties, J. Am. Chem. Soc. 127 (2005) 10982–10990, https://doi.org/10.1021/ja051240n.
[44] I. Nakamura, N. Negishi, S. Kutsuna, T. Ihara, S. Sugihara, K.J. Takeuchi, Role of oxygen vacancy in the plasma-treated $TiO_2$ photocatalyst with visible light activity for NO removal, J. Mol. Catal. A 161 (2000) 205–212. https://doi.org/10.1016/S1381-1169(00)00362-9.
[45] M. Li, X. He, Y. Zeng, M. Chen, Z. Zhang, H. Yang, P. Fang, X. Lu, Y. Tong, Solar-microbial hybrid device based on oxygen-deficient niobium pentoxide anodes for sustainable hydrogen production, Chem. Sci. 6 (2015) 6799. https://doi.org/10.1039/C5SC03249K.
[46] Y. Wang, Y. Chen, Y. Zuo, F. Wang, J. Yao, B. Li, S. Kang, X. Li, L. Cui, Hierarchically mesostructured $TiO_2$/graphitic carbon composite as a new efficient photocatalyst for the reduction of $CO_2$ under simulated solar irradiation, Catal. Sci. Technol. 3 (2013) 3286-3291. https://doi.org/10.1039/C3CY00524K.
[47] Z. He, L. Wen, D. Wang, Y. Xue, Q. Lu, C. Wu, J. Chen, S. Song, Photocatalytic reduction of $CO_2$ in aqueous solution on surface-fluorinated anatase $TiO_2$ nanosheets with exposed {001} facets, Energy Fuels. 28 (2014) 3982-3993. https://doi.org/10.1021/ef500648k.
[48] Y. Li, C. Wang, M. Song, D. Li, X. Zhang, Y. Liu, $TiO_{2-x}/CoO_x$ photocatalyst sparkles in photothermocatalytic reduction of $CO_2$ with $H_2O$ steam, Appl. Catal. B 243 (2019) 760–770. https://doi.org/10.1016/j.apcatb.2018.11.022.
[49] J. Jiao, Y. Wei, Y. Zhao, Z. Zhao, A. Duan, J. Liu, Y. Pang, J. Li, G. Jiang, Y. Wang, AuPd/3DOM-$TiO_2$ catalysts for photocatalytic reduction of $CO_2$: High efficient separation of photogenerated charge carriers, Appl. Catal. B 209 (2017) 228-239. https://doi.org/10.1016/j.apcatb.2017.02.076.
[50] E.A. Kozlova, M.N. Lyulyukin, D.V. Markovskaya, D.S. Selishchev, S.V. Cherepanova, D.V. Kozlov, Synthesis of $Cd_{1-x}Zn_xS$ photocatalysts for gas-phase $CO_2$ reduction under visible light, Photochem. Photobiol. Sci. 18 (2019) 871-877. https://doi.org/10.1039/C8PP00332G.
[51] L. Ye, H. Wang, X. Jin, Y. Su, D. Wang, H. Xie, X. Liu, X. Liu, Synthesis of olive-green fewlayered BiOI for efficient photoreduction of $CO_2$ into solar fuels under visible/near-infrared light, Sol. Energy Mater. Sol. Cells 144 (2016) 732-739. https://doi.org/10.1016/j.solmat.2015.10.022.
[52] T. Ye, W. Huang, L. Zeng, M. Li, J. Shi, $CeO_{2-x}$ platelet from monometallic cerium layered double hydroxides and its photocatalytic reduction of $CO_2$, Appl. Catal. B 210 (2017) 141-148. https://doi.org/10.1016/j.apcatb.2017.11.011.
[53] E. Pastor, F. Pesci, A. Reynal, A. Handoko, M. Guo, X. An, A. Cowan, D. Klug, J. Durrant, J. Tang, Interfacial charge separation in $Cu_2O/RuO_x$ as a visible light driven $CO_2$ reduction catalyst, Phys. Chem. Chem. Phys. 16 (2014) 5922-5926. https://doi.org/ 10.1039/C4CP00102H.
[54] K. Wang, J. Lu, Y. Lu, C.H. Lau, Y. Zheng, X. Fan, Unravelling the C-C coupling in $CO_2$ photocatalytic reduction with $H_2O$ on Au/$TiO_{2-x}$: combination of plasmonic excitation and oxygen vacancy, Appl. Catal. B 292 (2021), 120147. https://doi. org/10.1016/j.apcatb.2021.120147.
[55] L.Y. Lin, S. Kavadiya, X. He, W.N. Wang, B.B. Karakocak, Y.C. Lin, M.Y. Berezin, P. Biswas, Engineering stable Pt nanoparticles and oxygen vacancies on defective $TiO_2$ via

introducing strong electronic metal-support interaction for efficient $CO_2$ photoreduction, Chem. Eng. J. 389 (2020), 123450. https://doi.org/10.1016/j. cej.2019.123450.
[56] S. Wang, X. Han, Y. Zhang, N. Tian, T. Ma, H. Huang, Inside-and-out semiconductor engineering for $CO_2$ photoreduction: from recent advances to new trends, Small Struct. 2 (2021) 2000061. https://doi.org/10.1002/sstr.202000061.
[57] S. Akrami, P. Edalati, K. Edalati, M. Fuji, High-entropy ceramics: review of principles, production and applications, Mater. Sci. Eng. R 146 (2021) 100644. https://doi.org/10.1016/j.mser.2021.100644.
[58] F.Y. Liu, Y.M. Dai, F.H. Chen, C.C. Chen, Lead bismuth oxybromide/graphene oxide: synthesis, characterization, and photocatalytic activity for removal of carbon dioxide, crystal violet dye, and 2-hydroxybenzoic acid, J. Colloid. Interface Sci. 562 (2020) 112-124. https://doi.org/10.1016/j.jcis.2019.12.006
[59] H.L. Chen, F.Y. Liu, X. Xiao, J. Hu, B. Gao, D. Zou, C.C. Chen, Visible-light-driven photocatalysis of carbon dioxide and organic pollutants by $MFeO_2$ (M = Li, Na, or K), J. Colloid. Interface Sci. 601 (2021) 758-772. https://doi.org/10.1016/j.jcis.2021.05.156.